\documentstyle[preprint,aps,pra]{revtex}
\begin{document}
\draft
\title{A New Time-Scale for Tunneling}
\author{E. Eisenberg and Y. Ashkenazy}
\address{
Department of Physics, Bar-Ilan University, Ramat-Gan 52900, Israel}
\date{\today }
\maketitle

\begin{abstract}
We study the tunneling through an oscillating delta barrier.
Using time periodicity of the model, the time-dependent Schr\"odinger
equation is reduced to a simple but infinite matrix equation. Employing 
Toeplitz matrices methods, the infinite matrix is replaces by a 
$3\times 3$ matrix, allowing an analytical solution. Looking at the
frequency dependence of the transmission amplitudes, one observes a 
new time scale which dominates the tunneling dynamics. This time
scale differs from the one previously introduced by B\"uttiker and 
Landauer. The relation between these two is discussed.
\end{abstract}


\narrowtext
\bigskip\bigskip
\section{Introduction} \label{s1}
The question regarding the actual time spent during tunneling 
through a barrier has been in discussion for over half a century
\cite{old,wig,bl,hs,or,lmr}. The introduction of high-speed 
tunneling-based semiconductor devices in recent years \cite{exp}
has brought new urgency to the problem. However, in 
spite of the long history of this subject it still remains 
controversial.

Most of the opinions on this matter can be classified into three 
categories. The most apparent way to define and calculate tunneling 
times is through the dynamics of the wave-packet. Some suggestions
use the stationary phase approximation and obtained the time the peak 
of the wave packet spent in the barrier region \cite{wig}. In a more 
recent versions of this approach, the peak is replaced by the centroid 
of the packet \cite{h87}. Nevertheless, this approach is problematic. 
It has been shown \cite{lm}
by an explicit example that a tunneling situation can be set up such 
that the peak or centroid of the transmitted packed emerges before the
peak or centroid of the incident packet has even arrived to the 
barrier, leading to negative time and demonstrating the lack of causal 
relationship.

A more sophisticated way involves the introduction of a physical clock
which is used to determine the time elapsed during tunneling. One can
look at either the effect of the clock on the tunneling particle or
at the clock variable itself. Mainly, three versions of clocks have 
been proposed and investigated in the literature. The first clock
approach studies the precession of the spin of the tunneling particle
due to a uniform infinitesimal magnetic field confined to the barrier 
region \cite{baz}. It was later noted that in addition to the 
precession in the plane perpendicular to the magnetic field, the 
spin-up component in the magnetic field direction tunnels preferentially 
due to the Zeeman splitting \cite{b83}. These two processes caused by 
the field can be used as physical clocks. One gets two different time 
scales from these two. The two time scales are equal to the real and 
imaginary parts of the complex time obtained using the Feynman approach
\cite{sb}. A second clock approach uses an incident wave composed of 
two interfering waves \cite{bl85}.
If these two are of opposite spin direction, the total incident
particle flux is uniform in time with an oscillatory spin.

B\"uttiker and Landauer (BL) studied tunneling through a rectangular
barrier with a small oscillating component added to the height
\cite{bl}. The
incident particles with energy $E$ can absorb or emit modulation 
quanta $\hbar\Omega$ during tunneling, leading to the appearance of 
sidebands with energies $E\pm n\hbar\Omega$ and corresponding 
intensities $I_{\pm n}$. Looking at the $\Omega$ dependence of these 
quantities, BL determined the critical frequency as that in which the 
transition between the behavior at the two limits of quasi-static barrier 
and the 
average barrier occurs. The inverse of this frequency is an indication
of the time-scale of the tunneling process.

Third, there are many who object the question in the first place
\cite{dm}. It is often said that the concept of tunneling time is not 
well-defined in the context of quantum mechanics, and does not 
correspond to any observable. This is related to the absence of a 
time-operator in quantum mechanics. This problem can be bypassed using 
an operator which measures whether the particle is in the barrier or not, 
and then (in time-dependent problems) averaging over time. The result, 
divided by the incident flux, is termed the ``dwell time'' \cite{b83}. 
Nevertheless, this solution is far from being satisfactory, since taking 
expectation values, one can not distinguish between the transmitted and 
reflected particles, and thus the time obtained is just an average over 
two (possibly different) times related to these two processes. This 
problem follows from the absence of a well-defined history of a quantum 
particle \cite{s}. As far as the particle was not measured in the barrier, 
it was not there at all and thus formally the time spent in the barrier 
is zero \cite{larry}.

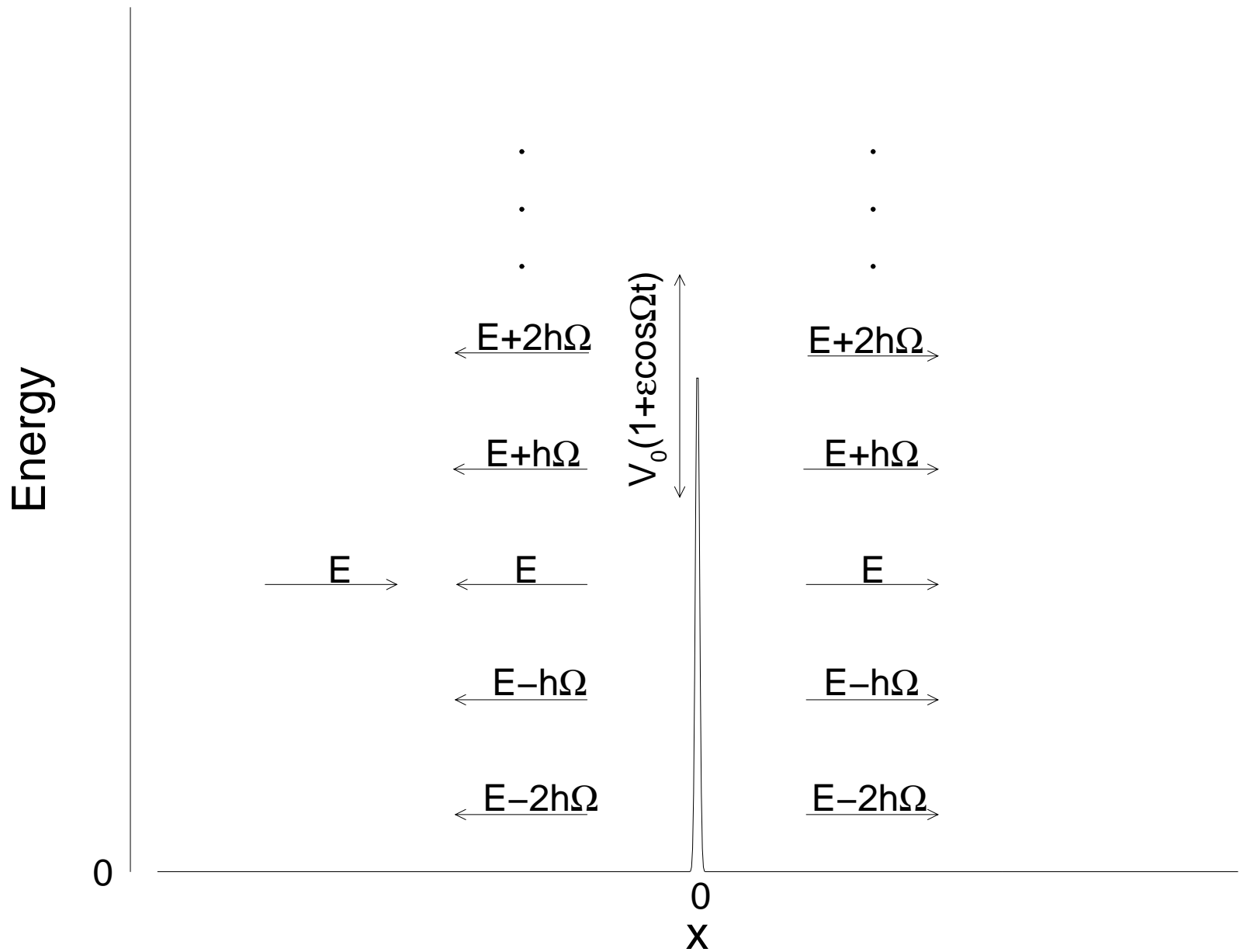
\begin{figure}[tbp]
\caption{Oscillating delta function barrier.} 
\label{fig1} 
\end{figure}

In this paper we do not want to take a stand in this debate, but 
rather to point out that even if one argues that the {\it actual} time
spent inside the barrier region is not defined or zero, it is agreed
that the {\it wave function} spent a finite time inside the barrier.
The question about the time-scales related to the wave-function 
transition through the barrier is at least as important as the (maybe
more fundamental) former one. After all, the quantum dynamics is 
determined by the wave-function and thus many questions in which time 
scales are relevant, are really not related to the {\it actual} time 
as measured experimentally, but rather at the time in which the 
wave-packet - describing the particle in times it is {\it not measured} 
- was in the regime in interest. This approach is motivated by the
physical clocks approach. Again, even if the application of the 
proposed experiments to the `strong' question of tunneling time is 
arguable, one can't deny the need for tunneling time scales while 
considering the process in a time-dependent environment such as 
considered in the BL approach. In what follows we use the same 
approach and consider the tunneling process through an oscillating 
delta barrier
\begin{equation}
\label{e0}
V(x,t)=V_0\delta(x)(1+\varepsilon\cos(\Omega t)).
\end{equation}
As in the BL case, the tunneling particle can absorb or emit any number
of energy quanta $\hbar\Omega$ while being transmitted or reflected
(see Fig. \ref{fig1}).
Using the time periodicity of the Hamiltonian, the Schr\"odinger 
equation is reduced to the inversion of an infinite tridiagonal
matrix. In practice, the matrix can be truncated very efficiently and
reliable numerical solutions are easily obtained. In the large quantum 
numbers regime the three-diagonals elements become almost constant, and
the matrix is approximately a Toeplitz matrix \cite{t1,t2}. In this case 
one can reduce the problem to the inversion of a $3\times 3$ matrix, and 
obtain analytical expressions.

Following Ref. \cite{bl,bl85}, we study the frequency dependence of the 
transmission amplitudes in the limit of small oscillation amplitude 
$\varepsilon$. We show that there exists a time scale $\tau$, 
such that the side-band asymmetry (studied by BL) is a function of
$\Omega\tau$. This time scale is not the same as the one defined
by BL which vanishes in the delta barrier limit. This indicates
the possibility of more than one time-scale for the tunneling process.

\section{Model} \label{s2}
\subsection{Analytical Solution}

We consider a time periodic potential (\ref{e0})
where $V_0$ is the barrier strength, $\varepsilon$ ($0 < \varepsilon < 1$)
is the modulation strength amplitude and $\Omega$ the barrier frequency.

We first note that since the problem is periodic the general solution
of the wave equation is a superposition of eigenstates of the Floquet
operator, i.e., the operator which shifts the solution in time by one
period. Thus we can solve for each such eigenstate separately. Since
the frequency of the incoming part of the asymptotic behavior of the 
solution is determined by the incident energy, this energy fixes the 
corresponding eigenvalue to be $e^{-iE_iT/\hbar}$, where $E_i$ stands 
for the incident energy. The outgoing parts
of the asymptotic solution correspond to the transmitted and reflected 
particles, and their frequencies must match the same eigenvalue, i.e.,
$e^{-iE_f T/\hbar}= e^{-iE_iT/\hbar}$, where $E_f$ is the final energy, after
transmission or reflection. We thus conclude that the frequency of all the 
asymptotic outgoing waves should differ from the initial frequency by an 
integer multiple of $\Omega$, i.e., the energy change is an integer number
of energy quanta $\hbar\Omega$.

The corresponding Schr\"odinger equation is 
\begin{equation}
i \hbar \frac{\partial}{\partial t} \psi = (- \frac{\hbar ^2}{2 m} 
\frac{\partial ^2}{\partial x^2} + V(x,t) ) \psi. \label{e2}
\end{equation}
On both sides of the barrier the solution of Eq. (\ref{e2}) is just a 
free particle solution. We thus separate the solution, and write it on both  
sides as a superposition of free particle functions, satisfying the
scattering boundary conditions, and having appropriate frequencies:
\begin{equation}
\psi(x,t) = 
  \left\{
    \begin{array}{ll}
      \psi_l(x,t) = e^{(ik_0x-i \omega_0 t)}+\sum_n
	 r_n e^{(-ik_n x-i \omega_n t)} & x < 0 \\
      \psi_r(x,t) = \sum_n
	 t_n e^{(ik_n x-i \omega_n t)} & x > 0
    \end{array}
  \right.    \label{e3}
\end{equation}
where $r_n$ and $t_n$ are the reflection and transmission coefficients,
$\omega_n=(E+n\hbar\Omega)/\hbar$, $k_n=\sqrt{2m\omega_n/\hbar}$ and the
summations are over every integer $n$.
For negative $\omega_n$, $k_n$ become imaginary, $k_n=i\kappa_n$. The 
corresponding wave-functions can not be extended to the whole space, 
but still can contribute to the wave-function on one side. One picks the 
decaying solution to ensure integrability.

Continuity of the wave function requires $\psi_l(0,t)=\psi_r(0,t)$, 
implying the following relations,
\begin{equation}
  \begin{array}{ll}
    r_n = t_n, & n\ne 0\\
    1+r_0=t_0, & n=0.
  \end{array}  \label{e4}
\end{equation}
Integrating Eq. (\ref{e2}) over the barrier region one obtains
a tridiagonal matrix equation :
\begin{equation}
(2ik_n/B-1)r_n-\frac{\varepsilon}{2}(r_{n+1}+r_{n-1}) = \delta_{0,n}
+ \frac{\varepsilon}{2}(\delta_{0,n+1}+\delta_{0,n-1}) , \label{e5}
\end{equation}
where $B=2mV_0/\hbar^2$.

In order to solve Eq. (\ref{e5}) we need to invert an infinite matrix. 
Practically, it is enough to truncate the matrix after about a dozen rows 
around the center ($n=0$), since the far $r_n$'s approach zero quickly
(we denote this solution by FS, for full solution). Moreover, 
it is easy to show that for $|n|\gg 1$ the matrix elements are approximately
constant. In this case we can treat our matrix as a Toeplitz matrix
\cite{t1}, and assume that the solution decays exponentially. The solution 
is then obtained by inversion of a $3\times 3$ matrix (we call this 
solution TS for Toeplitz solution).

Let us now focus on the TS. The method of reduction of an infinite
approximate Toeplitz matrix into a small finite one has been described
elsewhere \cite{t2}. We thus only briefly sketch the solution. 
We have to substitute the exponential correction to
the $3 \times 3$ matrix; the new matrix equation is :
\begin{equation}
 \left(
  \begin{array}{ccc}
    - \frac{\varepsilon}{2} e^{-\theta_-} + \frac{2i}{B}k_{-1} -1 &
	- \frac{\varepsilon}{2} & 0 \\
    - \frac{\varepsilon}{2} & \frac{2i}{B}k_{0} -1 & 
	- \frac{\varepsilon}{2} \\
    0 & - \frac{\varepsilon}{2} & 
	- \frac{\varepsilon}{2} e^{-\theta_+} + \frac{2i}{B}k_{1} -1
  \end{array}
 \right)
 \left(
  \begin{array}{c}
    r_{-1} \\
    r_0 \\
    r_1
  \end{array}
 \right)
=
 \left(
  \begin{array}{c}
    \frac{\varepsilon}{2} \\
    1 \\
    \frac{\varepsilon}{2}
  \end{array}
 \right),  \label{e6}
\end{equation}
where $\theta_-$ and $\theta_+$ are complex exponents characterizing the
(exponential) decay of the $t_n$'s on both sides of the central energy band,
$n=0$), ($\Re(\theta_-), \Re(\theta_+) > 0$). These exponents can be found 
by assuming that $r_n = r_{-1} \exp ((n+1) \theta_-)$ for $n \le -1$ and 
$r_n = r_{1} \exp (-(n-1) \theta_+)$ for $n \geq 1$, and substituting into 
Eq. (\ref{e5}).
We discuss three regimes : (a) $E>2\hbar\Omega$ ($k_{-1}$ 
and $k_{-2}$ are positive), (b) $\hbar\Omega<E<2\hbar\Omega$ ($k_{-1}$ positive
and $k_{-2}$ imaginary), and (c) $E<\hbar\Omega$ ($k_{-1}$ 
and $k_{-2}$ are imaginary). 
In the first regime the solution is :
\begin{equation}
  \begin{array}{c}
    \cos (\Im(\theta_\pm)) = \frac{\sqrt{2}}{2 \varepsilon} \left( 1+ 
       \varepsilon^2+\frac{4}{B^2}k_{\mp 2}^2-\sqrt{(1-\varepsilon^2)^2+
	\frac{16}{B^4}k_{\mp 2}^4+\frac{8}{B^2}(1+\varepsilon^2)k_{\mp 2}^2} 
	\right)^{\frac{1}{2}}   \\
    \cosh (\Re(\theta_\pm)) = -\frac{1}{\varepsilon} 
	\cos (\Im(\theta_\pm)) ^ {-1} 
  \end{array}  \label{e7}
\end{equation}
while similar solutions can obtained for cases (b) and (c).

It is clear from Eq. (\ref{e7}) that as $\varepsilon$ becomes smaller, the
correspondence between FS and TS improves. Since we are interested in the 
limit $\varepsilon\to 0$, we regard the TS as an exact one. 
Fig. \ref{fig2} compares the FS and TS in all three cases mentioned before 
(a-c) when $\varepsilon = 1$, which is the worst case. There is a good 
agreement for the three central coefficients even in the extreme 
case of $E \ll \frac{2m}{\hbar^2}V_0^2$. Significant relative deviations
are obtained only for the levels whose population is exponentially small.
Much better correspondence is achieved for smaller $\varepsilon$, see Fig.
2d. Since in what follows we use only these three 
coefficients, this figure confirms the accuracy of the Toeplitz method 
for the following derivation. 

\subsection{Numerical Results}

We verify the above derivation by solving numerically the dynamical time 
dependent Schr\"odinger equation and comparing the results to the TS.
In order to simplify the numerical calculation the delta function barrier 
is placed in the center of an infinite well. The algorithm 
we use is as follows. First, we find the eigenfunction and eigenvalues 
of the static problem, for each time. Once these are given, the numerical 
calculation becomes much simpler using adiabatic perturbation theory, 
leading to a set of first order differential equations. Due to symmetry,
half of the matrix elements vanish. In the case of the delta function the odd 
eigenfunctions are time independent, thus the effective number of equations is 
reduced \footnote{The number of levels taken in the computation should include
at least the first few energy bands $\langle E\rangle\pm n\hbar\Omega$.}. 
The numerical solution is performed using the adaptive time 
step fifth-order Cash-Karp Runge-Kutta method \cite{nr}.
The initial wave function is a Gaussian wave packet located in the center of 
the left side of the barrier, moving
toward the oscillating barrier. The wave packet collides with the barrier and 
splits to reflecting and transmitted parts. We measure the probability to be
on the right side of the barrier after a collision.

The same probability is calculated using FS in the following way.
We average the transmitted probability current given by FS over 
the energy, using the weights obtained from the energy components of the 
initial wave packet taken above. The results of the analytic and numeric 
results are presented in Fig \ref{fig3}. The agreement between the two 
calculations is very good, confirming our derivation.

\section{Tunneling time Scales} \label{s3}

In order to investigate the tunneling time problem, one follows the strategy
explained above, inspired by BL, and studies the frequency dependence
of the transmission intensities in the limit of small $\varepsilon$.
In particular, one looks at the relative sideband asymmetry
\begin{equation}
F(\Omega)\equiv \frac{I_1-I_{-1}}{I_1+I_{-1}}, \label{e8}
\end{equation}
where $I_n=|t_n|^2$. BL show that for an opaque rectangular barrier, in the 
high frequency regime 
\begin{equation}
\label{th}
F(\Omega)=\tanh(\frac{m\kappa\Omega}{\hbar d})\equiv \tanh(\Omega\tau_{BL})
\end{equation}
where $\tau_{BL}$ is defined as the tunneling time. The frequency $\Omega
=\tau_{BL}^{-1}$ is the transition frequency, below which the behavior
is determined by the low frequency limit.

In the low frequency limit, BL obtained the relation
\begin{equation}
I_\pm=(\varepsilon V\tau/2\hbar^2)^2
\end{equation}
for an opaque finite barrier, whose height is $V$. Using this equation
as a {\it definition} of $\tau$, they got the following result for
a general barrier
\begin{equation}
\tau = \left( \frac{m}{\hbar\kappa^2} \right) \left[ \frac{(\kappa^2-k^2)^2 
\kappa^2 d^2
+ k_0^4(1+\kappa^2d^2) \sinh ^2 \kappa d + k_0^2\kappa d(\kappa^2-k^2)
\sinh 2 \kappa d}{4 k^2 \kappa^2 + k_0^4 \sinh ^2 \kappa d} \right]^{1/2}, 
\label{e9}
\end{equation}
where $\kappa=(2m/\hbar^2)^{1/2}\sqrt{V-E}$,
$k_0=(2mV/\hbar^2)^{1/2}$, $k=(2mE/\hbar^2)^{1/2}$ and $d$ is 
the barrier width. 
In the delta barrier limit this expression vanishes. The same result is 
obtained for the general rectangular barrier using the Larmor clock approach 
\cite{b83}.


In what follows we use the first approach of BL, namely, looking at the 
asymmetry, and show that for a delta barrier in the deep tunneling regime
$E\ll \frac{2m}{\hbar^2}V_0^2$, the frequency behavior depends on a new 
time scale. Using the above described Toeplitz method, we obtained for the
asymmetry function $F(\Omega)=-\Omega\tau_\delta$ where
\begin{equation}
\tau_\delta=\frac{2\hbar^3}{mV_0^2}.
\label{e12}
\end{equation}
This holds only in the regimes (a-b), i.e., when $E>\hbar\Omega$.
In the other regime the behavior is also a function of $\Omega\tau_\delta$,
$F(\Omega)=\sqrt{\Omega\tau_\delta-T_0}$, where $T_0$ is the tunneling
probability in the absence of oscillations.
Similar results for the model (\ref{e0}) were obtained by St{\o}vneng and Hauge
without using the Toeplitz approach \cite{new1}. 

One thus sees that following the approach taken by BL, one finds a new and 
different time scale than the one obtained via the Larmor clock, which was 
also given in BL's original work 
using the low frequency limit. The interpretation of this result
is not yet clear. We suggest that there are more than one time
scale in the tunneling through a general barrier. BL claimed that since
for the opaque barrier, in the high frequency regime one gets a clear
definition of $\tau=\tau_{BL}$ (Eq. (\ref{th})); this $\tau$ is {\it
the tunneling time} for the opaque barrier. Therefore, looking
at the low frequency regime and knowing the tunneling time, one can
extract a connection between the low frequency behavior and the 
tunneling time
\begin{equation}
\tau = \frac{2\hbar^2}{\varepsilon V}\sqrt{I_\pm}
\end{equation}
This relation was then used \cite{bl85} to define the tunneling 
time in the general
case. Now, if there is more that one time scale in the process,
there is no reason to take the time scale which dominates the
high frequency regime and connect it to the low frequency behavior
which may be dominated by a different time scale. Looking at a
delta barrier, in which the first time scale $\tau_{BL}$ vanishes and only
the second time scale plays a role, one indeed sees that both the high and 
low frequency regimes are dominated by the same (second) time scale 
$\tau_\delta$.


\acknowledgments{It is an honor to dedicate this work to our devoted
teacher Lawrence P. Horwitz. In the last few years we have had the opportunity
to benefit from his wide knowledge, many areas of interest
including almost every subject of modern fundamental physics,
and especially his kindhearted and willing assistance. These made our
first steps in the world of physics a personal as well as intellectual
experience.}

\begin{figure}[tbp]
\caption{Correspondence between FS (Full Solution) and TS (Toeplitz Solution).
$\hbar=2m=1$, $\Omega=\varepsilon=1$, $V_0=10$. (a) $E=2.5 > 2\Omega$; (b)
$\Omega<E=1.5<2\Omega$; (c) $E=0.5 < \Omega$; (d) Same as (a), but for 
$\varepsilon=0.5$.} 
\label{fig2} 
\end{figure} 
 
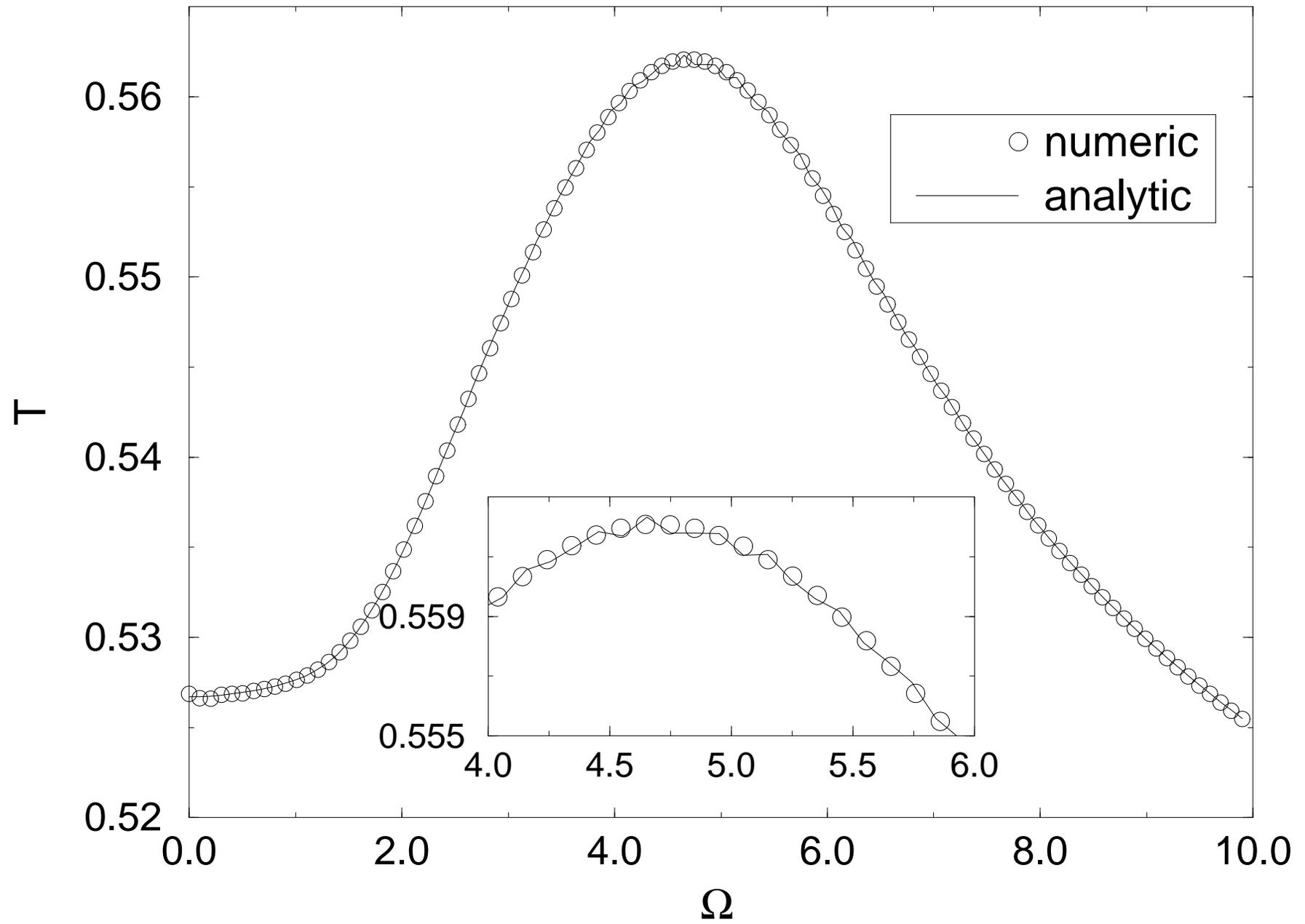
\begin{figure}[tbp]
\caption{Tunneling probability as a function of $\Omega$ - comparison of 
FS and dynamical computation. The units taken are 
such that $\hbar=2m=1$, $\langle E\rangle=5$, $V_0=5$, $\varepsilon=0.9$. 
The inset zooms on the maximum region}

\label{fig3} 
\end{figure} 
 
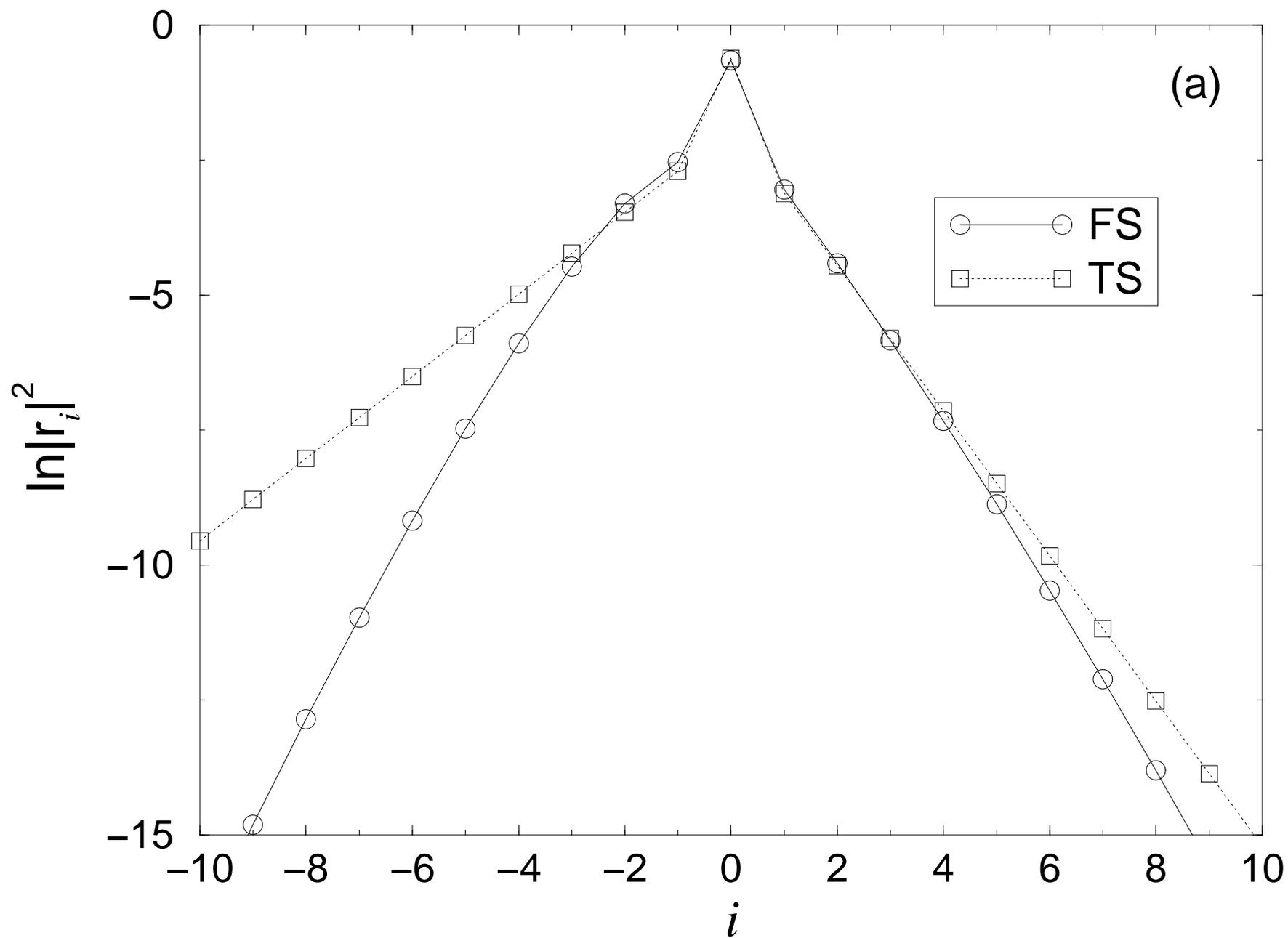
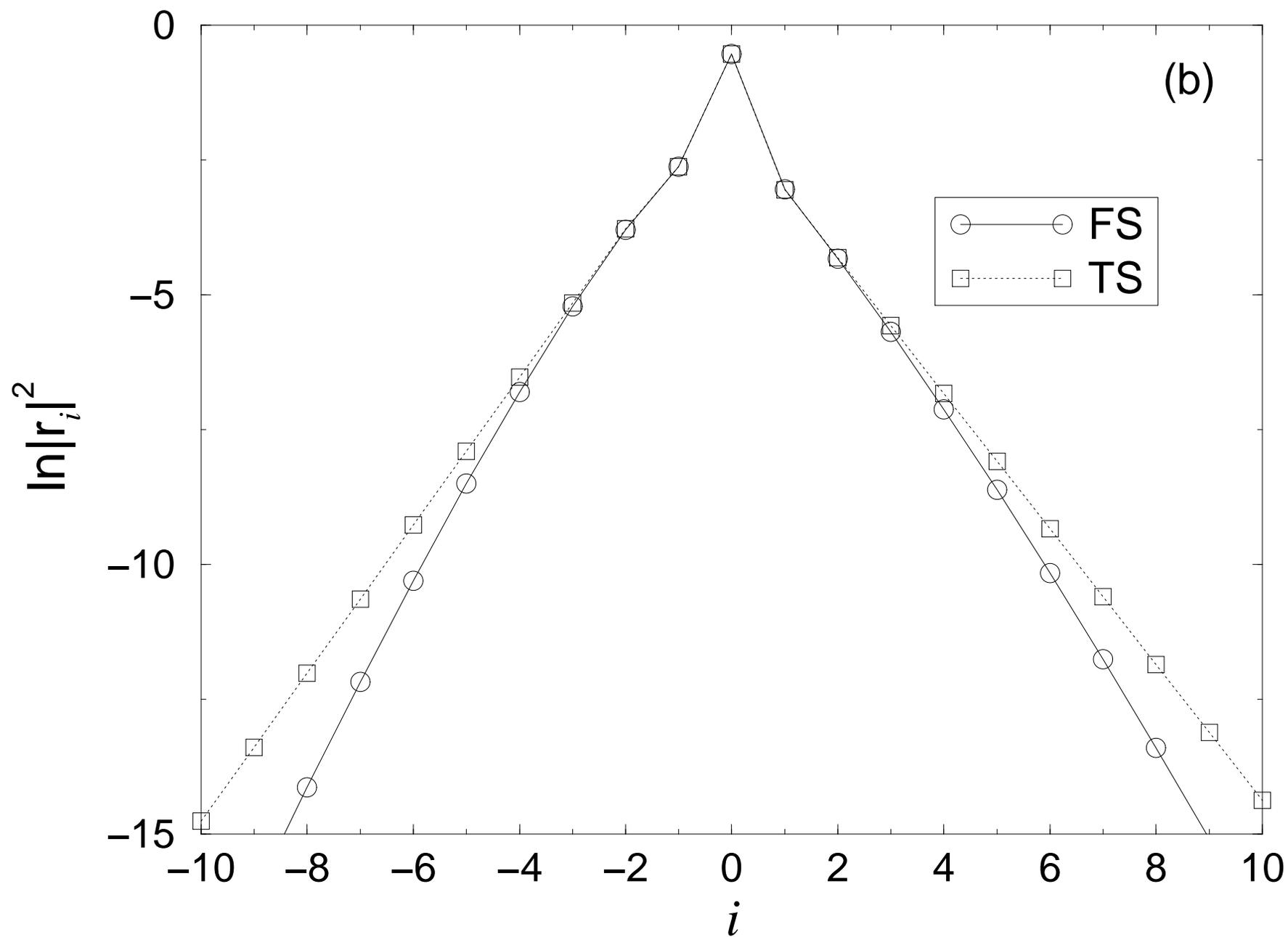
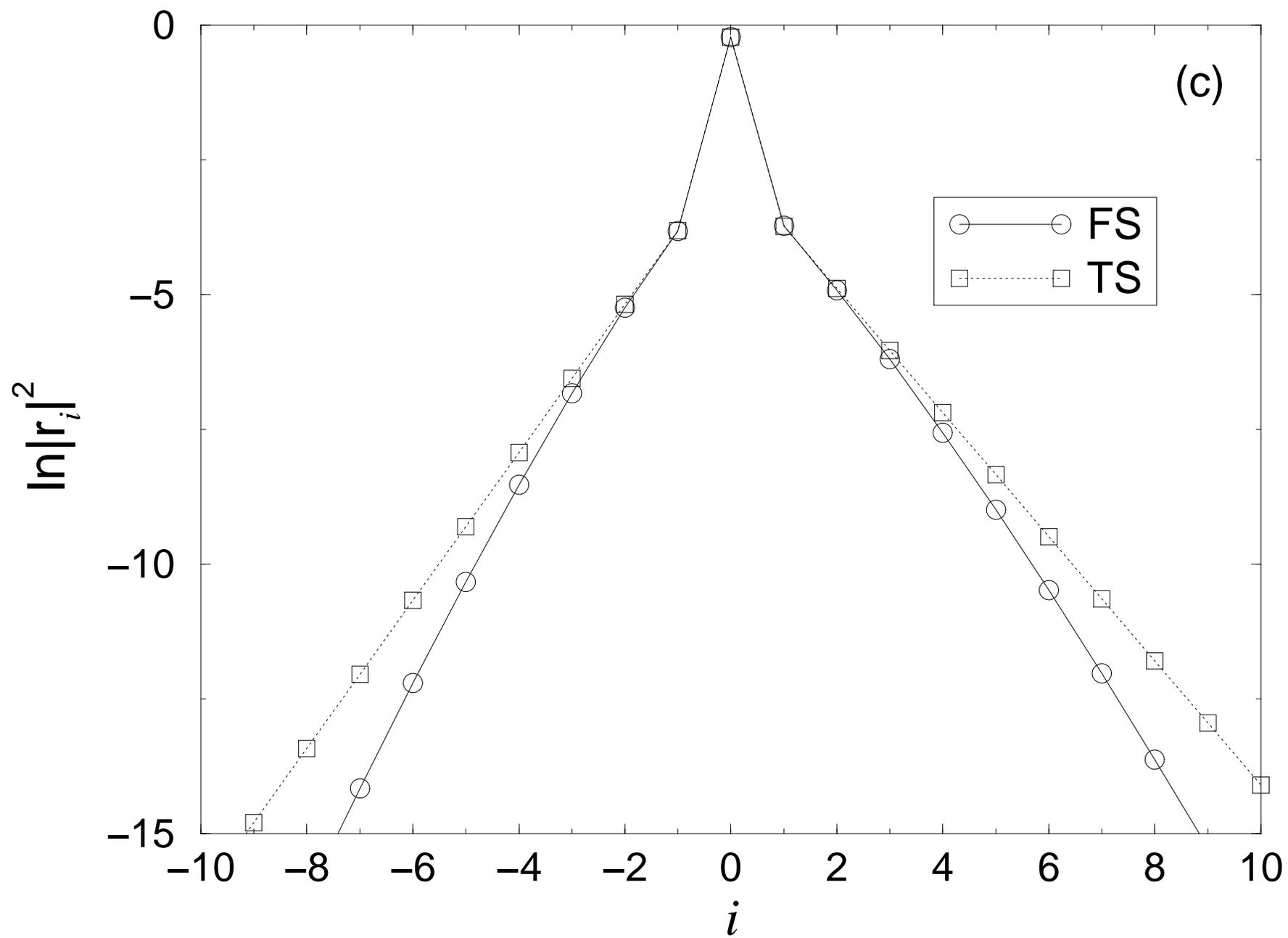
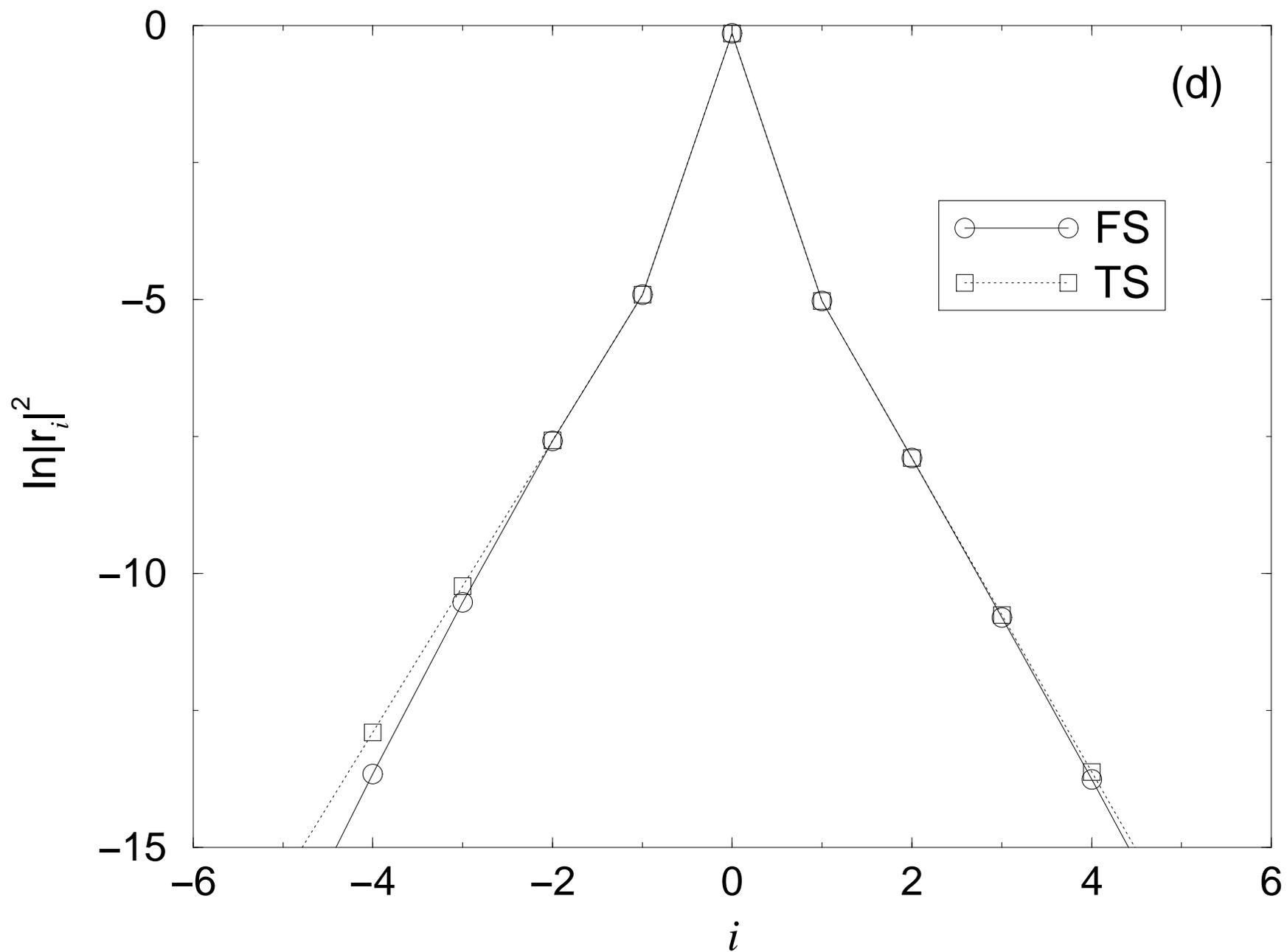
\end{document}